\documentclass{article}
\usepackage{spconf}
\usepackage{amsmath}
\usepackage{graphicx}
\usepackage{multirow}
\usepackage{makecell}
\usepackage{url}
\usepackage{times}
\usepackage{booktabs}
\usepackage{tabularx}
\usepackage{setspace}
\usepackage{afterpage}
\usepackage{cite}
\usepackage{url}
\usepackage{fancyhdr}
\newlength\mystoreparindent
\newlength\mystoreparskip
\newenvironment{mypara}[2]{%
    \setlength{\mystoreparindent}{\the\parindent}
    \setlength{\mystoreparskip}{\the\parskip}
    \setlength{\parindent}{#1}
    \setlength{\parskip}{#2}
    }{%
    \setlength{\parindent}{\mystoreparindent}
    \setlength{\parskip}{\mystoreparskip}
    }

\title{CST-former: transformer with Channel-Spectro-Temporal Attention for sound event localization and detection}

\name{Yusun Shul and Jung-Woo Choi\thanks{This work was supported by the National Research Council of Science \& Technology (NST) grant by the Korea government (MSIT) (No. CRC21011), and BK21 FOUR program through the National Research Foundation (NRF) funded by the Ministry of Education of Korea.}}
\address{School of Electrical Engineering, Korea Advanced Institute of Science and Technology\\ Daejeon, Republic of Korea \\\{shulys, jwoo\}@kaist.ac.kr}

\begin{document}
\maketitle

\begin{abstract}
Sound event localization and detection (SELD) is a task for the classification of sound events and the localization of direction of arrival (DoA) utilizing multichannel acoustic signals. Prior studies employ spectral and channel information as the embedding for temporal attention. However, this usage limits the deep neural network from extracting meaningful features from the spectral or spatial domains. Therefore, our investigation in this paper presents a novel framework termed the Channel-Spectro-Temporal Transformer (CST-former) that bolsters SELD performance through the independent application of attention mechanisms to distinct domains. The CST-former architecture employs distinct attention mechanisms to independently process channel, spectral, and temporal information. In addition, we propose an unfolded local embedding (ULE) technique for channel attention (CA) to generate informative embedding vectors including local spectral and temporal information. Empirical validation through experimentation on the 2022 and 2023 DCASE Challenge task3 datasets affirms the efficacy of employing attention mechanisms separated across each domain and the benefit of ULE, in enhancing SELD performance.

\end{abstract}
\section{Introduction}
\label{sec:intro}
Sound event localization and detection (SELD) refers to a multi-task for sound event detection (SED) and direction-of-arrival estimation (DoAE) in three-dimensional (3-D) space over time. The Detection and Classification of Acoustic Scenes and Events (DCASE) challenge task3 \cite{2018DCASE_baseline} introduced this task aiming to develop deep neural network (DNN)-based SELD models robust to unseen DoA values, reverberation, and low signal-to-noise-ratio (SNR) scenes. DCASE task3 provides polyphonic datasets including 13 types of audio classes and two types of multi-channel array signals: first-order-Ambisonics (FoA) and tetrahedral microphone array signals. 

To accomplish SED and DoAE tasks, models were constructed as multi-task networks, e.g., a single branch network producing multiple losses~\cite{2019overview_TASLP,2020Du_task3_1st,2020Park_task3_5th}. Nguyen et al.~\cite{2020Nguyen_task3_2nd} designed two modules dedicated to SED and DoAE, whose outputs are combined afterward to predict SED and DoA. 
In addition, soft-parameter sharing approaches were suggested to enable information exchange between SED- and DoAE-related features~\cite{einv2}. 
However, Shimada et al.~\cite{accdoa} introduced activity-coupled cartesian DoA (ACCDOA) that couples the outputs of SED and DoAE as the length and direction of a vector, enabling the design of the model as a single task. 
Recently, multi-ACCDOA~\cite{multiaccdoa} was introduced to overcome the limitation of ACCDOA in detecting in-class polyphonies through the class-wise auxiliary duplicated permutation invariant training (PIT).
Also, Kim et al.~\cite{ad-yolo} presented angular-distance-based multiple SELD output to secure robustness against unknown polyphonic events. 

The datasets for 2022 and 2023 DCASE challenge task3 are STARSS22~\cite{dcase2022baseline_workshop} and STARSS23~\cite{data_starss23} of which audio signals were recorded in real scenes with various room-impulse-responses (RIRs), unlike the synthetic audio datasets used for the challenge until 2021. Due to the limited number of real data, synthetic data (\emph{synth-set})~\cite{data_DCASE2022synth} generated by convolving single channel data FSD50K~\cite{FSD50K} and TAU-SRIRs~\cite{data_tau_srirDB} were provided. 
Recently, Wang et al.~\cite{ACS_TASLP} and Niu et al.~\cite{ResnetConformer} have demonstrated that proper augmentation and utilization of various external data are profitable to the SELD performance. 

\begin{figure*}[ht]
\centering
\centerline{\includegraphics[width=0.8\textwidth]{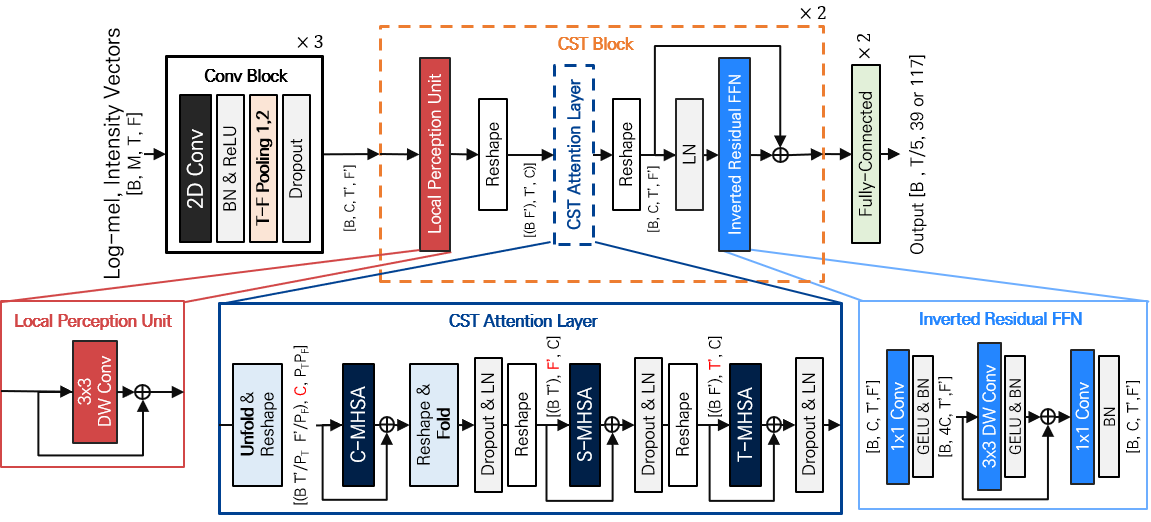}}
\caption{Architecture of Channel-Spectro-Temporal(CST)-former with unfolded local embedding (ULE) for channel attention. T-F pooling 1 and 2 indicate pooling at the front and middle layers, respectively.}
\label{fig:CST_former}
\end{figure*}

Nevertheless, advanced DNN architectures are expected to demonstrate the ability to encode spatial relations within given data, without excessive dependence on data augmentation techniques or the incorporation of supplementary external datasets. 
Most frequently used architectures such as ResNet-Conformer~\cite{ResnetConformer}, event independent network V2 (EINV2)~\cite{einv2}, and convolutional recurrent neural network (CRNN)~\cite{dcase2022baseline_workshop} in Table 1 employ temporal attention or gated recurrent unit (GRU) to analyze temporal context. These mechanisms use spectral and channel information as the embedding of the time sequence, which limits a DNN model to extract useful relations in spectral or spatial dimensions. Because research in other fields such as speech enhancement~\cite{TPARN,dasformer,DeftAn} and video classification~\cite{videot_dividedatten} has proved that applying independent attention mechanisms for different domains is crucial, we formerly introduced separated attentions for the temporal and spectral domains in form of divided spectro-temporal (DST) attention~\cite{DST_attention}. 
Although DST attention outperformed the baseline model of DCASE 2023 challenge, it has clear limitations in that the feed-forward network (FFN) is missing and channel information is utilized only as an embedding of spectral and temporal sequences. Therefore, in this work, we propose a Channel-Spectro-Temporal Transformer (CST-former) that employs distinct attention for all three domains: time, frequency, and channel. In addition, we introduce unfolded local embedding (ULE) operation for constructing useful channel attention (CA) embeddings and demonstrate its efficacy on SELD performance.

\begin{figure}
\centering
\centerline{\includegraphics[width=1.05\columnwidth]{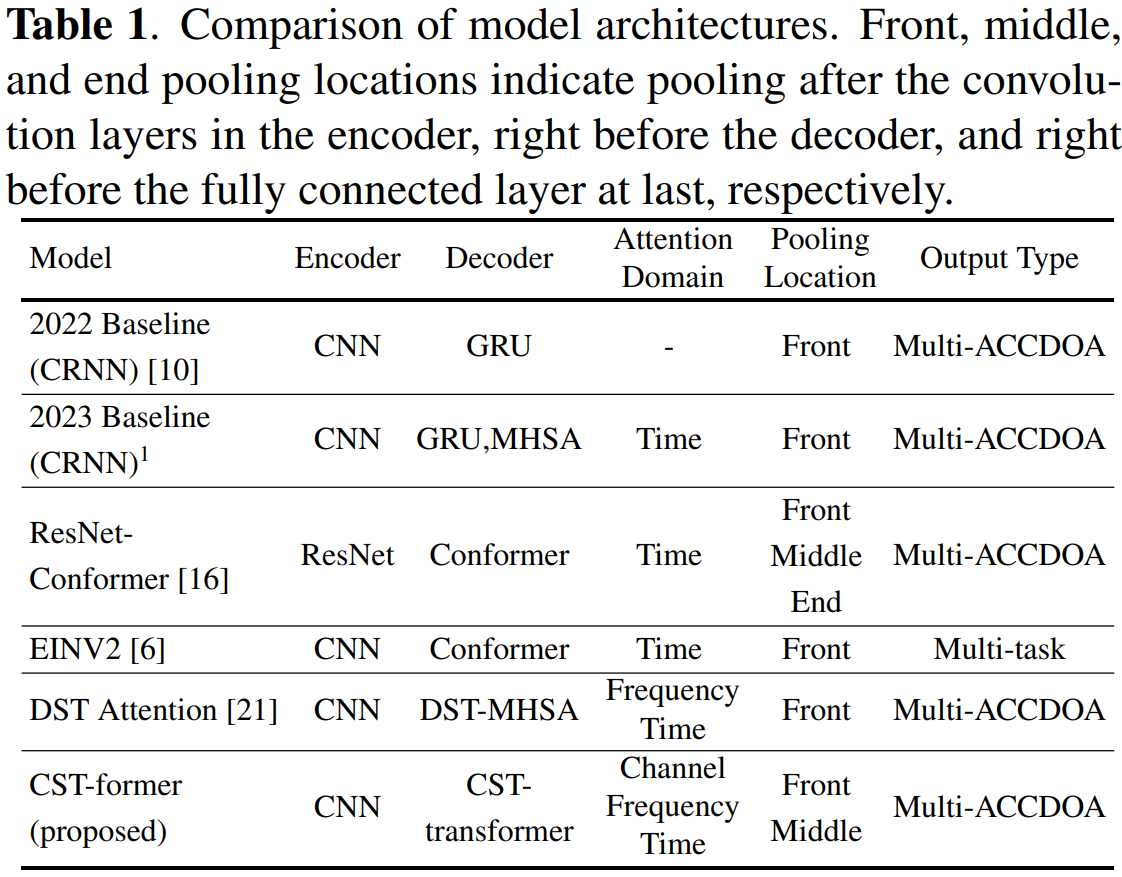}}
\label{tab: Models_table}
\end{figure}

\afterpage{\footnotetext[1]{https://github.com/sharathadavanne/seld-dcase2023}}

\section{Proposed Methods}
\label{sec:pro_meth}
\subsection{Channel-Spectro-Temporal Transformer}
\label{ssec:CST-former}
The overall architecture of the proposed CST-former is depicted in Fig.~\ref{fig:CST_former}. Log-mel spectrograms of FoA array signals and intensity vectors (IVs) are used as input, resulting in a total of $M=7$ input channels. The CST-former consists of three components, the conv block encoding the input data and applying time-frequency (T-F) pooling, the CST block for multi-domain attention, and fully connected (FC) layers for output estimation. 

\begin{mypara}{0pt}{3pt}
\textbf{\emph{Time-Frequency Pooling}} T-F pooling with different kernels is utilized in the conv block to match the temporal resolution of the target label and minimize the computational cost without sacrificing performance. 
Here, we consider two types of pooling; T-F pooling 1 (front) implements pooling with $(5,2)$, $(1,2)$, and $(1,1)$ kernels for three conv blocks, while T-F pooling 2 (middle) applies pooling in reverse order. These kernel sizes were derived from the analysis in \cite{DST_attention}, and both pooling layers produce feature outputs of the same size. 
\end{mypara}

\begin{figure*}[ht]
\centering
\centerline{\includegraphics[width=1\textwidth]{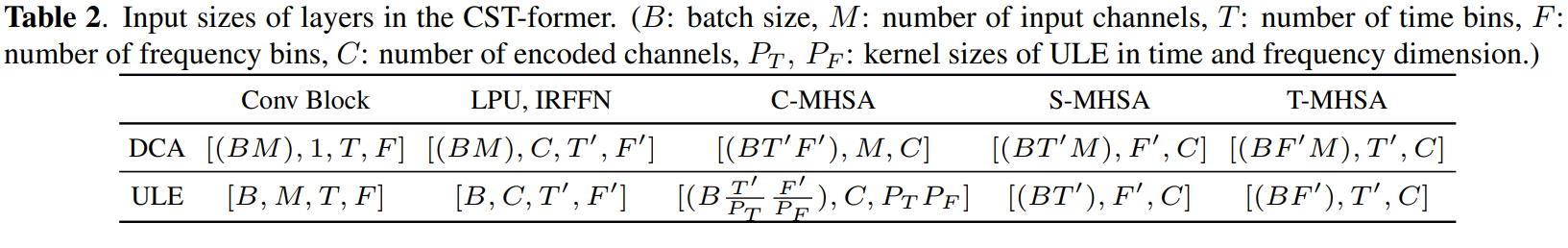}}
\label{tab: Input4CA}
\end{figure*}

\begin{mypara}{0pt}{3pt}
\textbf{\emph{CST Block}} The CST block contains the local perception unit (LPU), the CST attention layer, and inverted residual FFN (IRFFN). LPU and IRFFN are adopted from Convolution Meets Transformer (CMT) \cite{CMT_2022_CVPR} that integrates convolutional layers into vision transformer \cite{ViT} by substituting linear layered FFN with convolution layers. Local temporal and spectral information is extracted by $3\times3$ depth-wise convolution layer and residual in LPU. IRFFN consists of an expansion layer with a ratio of four, a depth-wise convolution with residual connection, and a projection layer. The usage of both LPU and IRFFN layers is indicated as CMT in this paper and CST-former includes the CMT. The CST attention in the CST block is explained in section~\ref{ssec:CST-attention}.
\end{mypara}

\subsection{Channel-Spectro-Temporal Attention}
\label{ssec:CST-attention}
The CST attention layer has three independent multi-head self-attentions (MHSAs) dedicated to three different domains: channel (C-MHSA), spectral (S-MHSA), and temporal (T-MHSA). In the previous work~\cite{DST_attention}, the $M$-channel input consisting of log-mel spectrograms and IVs is encoded into $C$-channel features by 2D convolution layers, and the encoded channel dimension is used as the embedding dimension for spectral and temporal attention. However, to implement CA across the channel dimension, an additional embedding dimension is required. In this paper, we propose two ways of embedding generation for CA: divided channel attention (DCA) and ULE.

\begin{mypara}{0pt}{3pt}
\textbf{\emph{Divided Channel Attention}} In the encoding stage, DCA simply treats the input channel dimension ($M$) as batch and outputs features of size $(BM,\, C,\, T^\prime,\, F^\prime)$ from the 2D convolutions in time and frequency dimensions. Afterward, the channel dimension of encoded features is used as the embedding dimension for all attention mechanisms. 
For subsequent attention, one dimension among time ($T^\prime$), frequency ($F^\prime$), and space ($M$) is selected as the sequence dimension, and the rest two dimensions are regarded as the batch dimension. Accordingly, the attention mechanism is similar to that of \cite{DST_attention}, except for the addition of CA. The input size of each block is presented in the second row of Table 2.
\end{mypara}

\begin{figure}[t!]
\centering
\centerline{\includegraphics[width=0.46\textwidth]{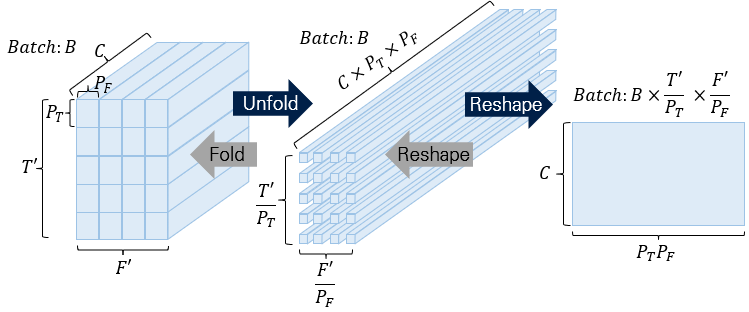}}
\caption{Unfolding and folding for the local embedding generation and reconstruction in channel attention.}
\label{fig:Unfold-Fold}
\end{figure}

\begin{mypara}{0pt}{3pt}
\textbf{\emph{Unfolded Local Embedding}} In ULE, the $M$-channel input is processed by the 2D conv block encoder, yielding the increased channel dimension from $M$ to $C$ and decreased T-F dimensions from $(T,\,F)$ to $(T^\prime,\,F^\prime)$. Instead of using the encoded channel as embedding for CA, ULE utilizes local temporal and spectral information obtained by unfolding the local T-F patch of size $(P_T, P_F)$, as shown in Fig.~\ref{fig:Unfold-Fold}.
For CA, individual patches are aligned along the batch dimension, resulting in the tensor size $(B T^\prime F^\prime/P_T P_F, C, P_T P_F)$. After being processed by CA, the output is reshaped to its original size by reversing the unfolding process. Since the data size is preserved before and after CA, the output from CA can be subsequently processed by temporal and spectral attention modules that use encoded channel dimension as embedding dimension. Since local T-F information is utilized in CA, ULE can consider spatial, spectral, and temporal information simultaneously. The input size for each layer of ULE is displayed in the third row of Table 2.
\end{mypara}

\section{Experiments and result analysis}
\label{sec:results}
\subsection{Implementation Details}
\label{ssec:implement_detail}
We used \emph{dev-set-train} of STARSS22 and STARSS23, as well as \emph{synth-set}, for training. \emph{dev-set-test} was used for evaluation. We trained the proposed model only on the FoA array signals. We applied the short-time Fourier transform (STFT) with a hop length of 0.02\,s and a window size of 0.04\,s on the audio signals of 24\,kHz sampling frequency. The spectrograms were transformed into log-mel scales using 64 mel filter banks. Also, IVs were calculated by multiplying the complex conjugation of the omnidirectional spectrograms with the other directional array spectrograms and taking the real part of the multiplication. These IVs were normalized using the total energy of the FoA signals. The total number of input channels was seven consisting of four original spectrograms and three intensity components. For comparison, multi-ACCDOA was used as the output format. The model was trained for 500 epochs with mean-squared error (MSE) loss and a batch size of 32. The Adam optimizer was employed and a tri-stage learning rate scheduler was used with an upper limit of 0.001, which comprises a ramp and a cosine scheduler. The input signals were segmented into 5\,s without overlapping. The number of conv blocks was three, and the number of filters for 2D convolution was 64. The number of CST blocks was two, and all MHSAs had eight multi-heads. $P_T$ and $P_F$ were set to ten and four, respectively. We evaluated our method with $\mathrm{SELD}_{score}$ from~\cite{2018DCASE_baseline} which is an average of four metrics: error rate (ER), macro-averaged F1-score (F), localization error (LE), and localization recall (LR). 

\subsection{Experimental Results}
\label{ssec:exper_res}
\subsubsection{Ablation studies for CST-former}
\label{sssec:Ablation}
Ablation studies were conducted to validate the efficiency of CA, CMT, and embedding generation methods (DCA and ULE) constituting the CST-former. The performance of ablated models trained and tested on the DCASE 2022 challenge dataset is presented in Table 3. 

Comparing between DCASE baselines, it is obvious that temporal attention incorporated in the DCASE 2023 baseline is helpful for the SELD task and improves the SELD performance by 6.3\% compared to the 2022 baseline. The separated T-F attention employed in DST attention is even more beneficial, lowering the SELD score by 9.0\%.
Furthermore, the integration of CA using DCA and ULE enhances the SELD performance by 11.1\% and 12.1\%, respectively. Notably, the performance of ULE surpasses that of DCA and DST attention, which demonstrates the importance of local T-F embedding in discovering relations between encoded channels.    

CMT is another important block providing extra performance gains of up to the 19.8\% SELD score compared to the 2022 baseline. Lastly, the change of T-F pooling location from Front to Middle contributes positively to the SELD performance, resulting in the final SELD score lowered by 22.1\% compared to the baseline model. This is a considerable amount of improvement achieved without using data augmentation techniques. 

\begin{figure}
\centering
\centerline{\includegraphics[width=1\columnwidth]{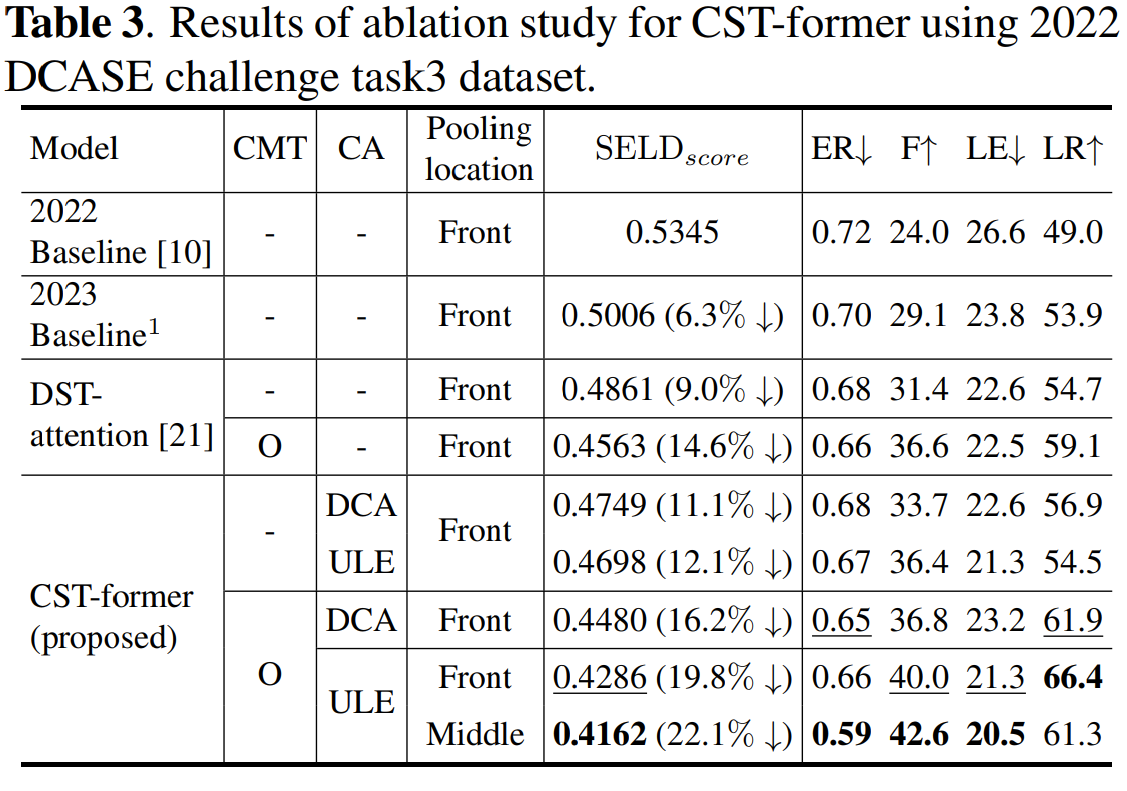}}
\label{tab:ablation}
\end{figure}

\begin{figure}
\centering
\centerline{\includegraphics[width=1\columnwidth]{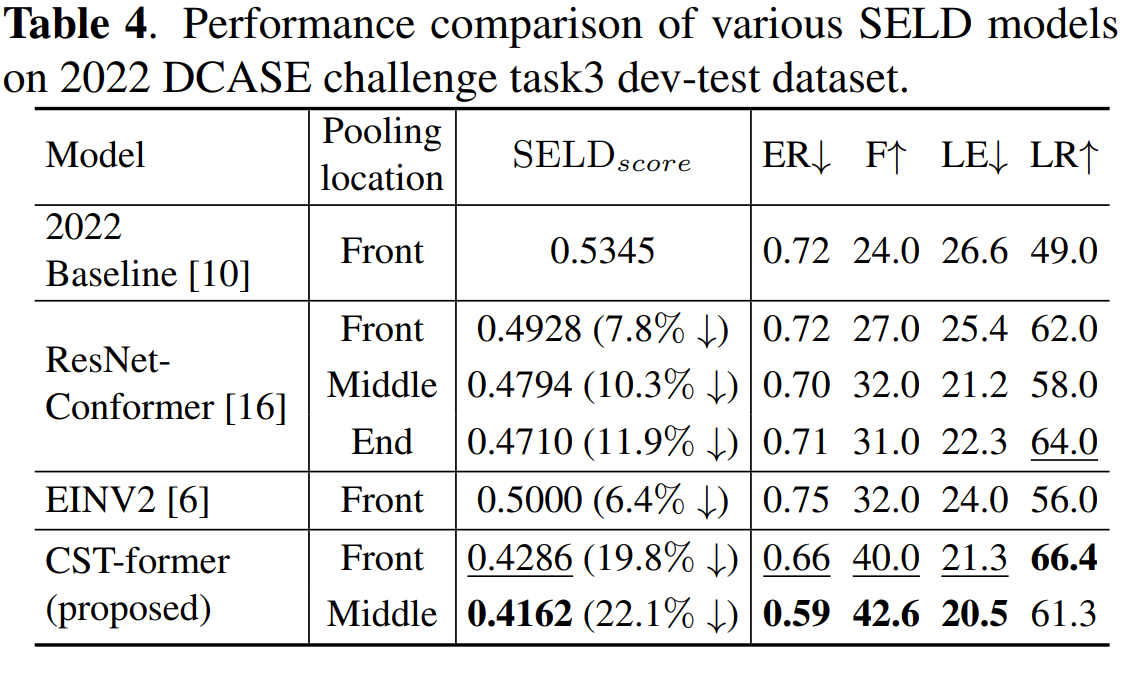}}
\label{tab:comparison_2022}
\end{figure}

\begin{figure}
\centering
\centerline{\includegraphics[width=1\columnwidth]{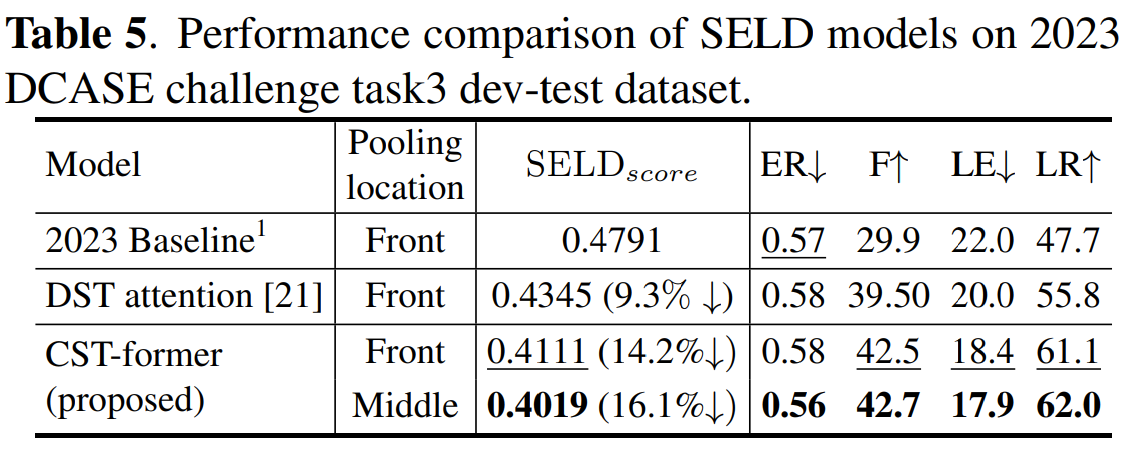}}
\label{tab:comparison_2023}
\end{figure}

\subsubsection{Verification with DCASE challenge task3 datasets}
\label{sssec:comparison}
For further validation of the proposed model, we compared the CST-former model to two well-established models, ResNet-Conformer and EINV2. First, the results presented in Table 4 show the performance of models trained and tested by the data from the 2022 DCASE challenge dataset. 
The CST-former demonstrates a remarkable achievement, showcasing a 22.1\% reduction in SELD score compared to the 2022 baseline. In contrast, ResNet-Conformer and EINV2 exhibit more modest improvements, with gains of only up to 11.9\% and 6.4\%, respectively. Furthermore, when trained and evaluated on the 2023 DCASE challenge dataset, as shown in Table 5, CST-former exhibits a substantial 16.1\% reduction in SELD score, surpassing the performance of the DST attention by 6.8\%. This empirical evidence underscores the efficacy of CA, CMT, and ULE under the limited number of training data without using data augmentation.

\section{Conclusion}
\label{sec:conc}
We propose the CST-former, a novel transformer-based architecture that employs distinct attention mechanisms for different domains for the SELD task. Furthermore, we introduce ULE as a key component for channel attention, utilizing the unfolded local temporal and spectral embedding as the channel embedding. Our experiments reveal that ULE outperforms DCA, DST-attention, and other top-tier models proposed for the SELD task. The CST-former consistently exhibits significant performance improvements on both the 2022 and 2023 DCASE challenge task3 datasets even without adopting any data augmentation techniques, establishing its effectiveness as a DNN architecture tailored for SELD with a limited size of training data in real scenes.

\vfill
\pagebreak

\begin{spacing}{0.94}
\bibliographystyle{IEEEbib}
\bibliography{strings}
\end{spacing}
\end{document}